\documentclass[12pt]{article}

\def\fnote#1#2{\begingroup\def\thefootnote{#1}\footnote{#2}\addtocounter{footnote}{-1}\endgroup}

\usepackage{amssymb}
\usepackage{makeidx}
\usepackage{epsf}
\usepackage{graphicx}        

\def\inbar{\vrule height1.5ex width.4pt depth0pt}
\def\IB{\relax{\rm I\kern-.18em B}}
\def\IC{\relax\,\hbox{$\inbar\kern-.3em{\rm C}$}}
\def\ID{\relax{\rm I\kern-.18em D}}
\def\IE{\relax{\rm I\kern-.18em E}}
\def\IF{\relax{\rm I\kern-.18em F}}
\def\IG{\relax\,\hbox{$\inbar\kern-.3em{\rm G}$}}
\def\IH{\relax{\rm I\kern-.18em H}}
\def\II{\relax{\rm I\kern-.18em I}}
\def\IK{\relax{\rm I\kern-.18em K}}
\def\IL{\relax{\rm I\kern-.18em L}}
\def\IM{\relax{\rm I\kern-.18em M}}
\def\IN{\relax{\rm I\kern-.18em N}}
\def\IO{\relax\,\hbox{$\inbar\kern-.3em{\rm O}$}}
\def\IP{\relax{\rm I\kern-.18em P}}
\def\IQ{\relax\,\hbox{$\inbar\kern-.3em{\rm Q}$}}
\def\IR{\relax{\rm I\kern-.18em R}}
\def\IT{\relax{\rm I\kern-.18em T}}
\def\ZZ{\relax{\sf Z\kern-.4em Z}}

\def\a{\alpha}         
    \def\k{\kappa}  \def\l{\lambda}

\def\cO{{\cal O}}

  \def\mathP{{\mathbb P}}

\def\fnote#1#2{\begingroup\def\thefootnote{#1}\footnote{#2}\addtocounter
{footnote}{-1}\endgroup}
\def\beq{\begin{equation}}
\def\eeq{\end{equation}}
\def\bea{\begin{eqnarray}}
\def\eea{\end{eqnarray}}

\let\nn=\nonumber

\def\notin{\ \hbox{{$\in$}\kern-.51em\hbox{/}}}

  \def\E1Fq{E_1/\IF_q}

\def\rmmeV{{\rm meV}}   
\def\rmmm{{\rm mm}}     \def\rmmod{{\rm mod}}
\def\rmnm{{\rm nm}}

\def\rmGeV{{\rm GeV}}

\def\rmTeV{{\rm TeV}}

\def\notdiv{{\relax{~|\kern-.35em /~}}}

\def\boxit#1{
\vbox{\hrule height1pt\hbox{\vrule width1pt\kern0.3cm
\vbox{\kern0.3cm\hbox{$\displaystyle#1$}\kern0.3cm}\kern0.3cm\vrule
width1pt}\hrule height1pt}}

\begin{document}
\parindent=0pt

\hfill \phantom{\bf DRAFT}

\hfill \phantom{\today}

\vskip 0.9truein

\parindent=0pt

{\bf APPLIED STRING THEORY}

\vskip .2truein

{\sc Rolf Schimmrigk\fnote{$\dagger$}{email: netahu@yahoo.com}}

\vskip .1truein

{\it Indiana University South Bend}
 \vskip .05truein
{\it 1700 Mishawaka Ave., South Bend, IN 46634}

\vskip  .3truein

\parindent=0pt
\pagenumbering{arabic}

\baselineskip=14pt
\parskip=.02truein

\tableofcontents

\vfill \eject

\parskip=.15truein

\baselineskip=19.5pt

\section{Introduction}
 The observation that the structure of string theory is rich
enough to include the standard model in rough outline is an old
one, starting with the early constructions of free field
constructions, orbifold theories, and in particular Calabi-Yau
compactifications in the late 1980s and early 1990s. At the time
 these constructions provided a large collection of different
vacua, with thousands of explicitly constructed Calabi-Yau
manifolds \cite{cls90}, and estimates of vast numbers of bosonic
models \cite{lls87}, each one associated with its own moduli
space. It was clear even then that it would be impossible to
systematically search this string vacua landscape. This, however,
is not a fundamental problem. Adopting the point of view that any
physical theory has to describe not only our universe, but all
possible consistent universes, leads to the obvious strategy of
using some phenomenological input to select viable models among
the ocean of models that obviously do not describe physics as we
know it.

Beyond the generic predictions made by all string models, such as
the existence of additional dimensions, and new particles, such as
the dilaton, a detailed phenomenological analysis of a few
three-generation Calab-Yau manifolds \cite{y85, s87}, lead to very
specific phenomenological features. The literature on this first
phase of string phenomenology is vast and it would be quite
impossible to summarize even these early developments in a short
review.

A venerable explosion of papers took place after Polchinski's
introduction of D-branes into the string theoretic landscape. A
great many attempts were made to generalize the notion of a brane,
and to incorporate them into ad-hoc models, purely
phenomenological constructions that aimed at scenarios not
necessarily embedded in a complete fundamental theory. The
possibility provided by branes of restricting physical modes to
lower dimensional submanifolds embedded in a higher dimensional
spacetime was used first in a class of models based on the
simplest possible toroidal framework in which all of the extra
dimensions are of equal size. Refinements of this first step have
opened a vast playground in which such ad hoc models were further
explored, making different assumptions about which fields are
confined to the low-dimensional brane, and which fields are
allowed to propagate in the bulk. One of the most interesting
consequences of this exploration in the late 1990s was the
observation, first made in \cite{add98, add99}, that if the
compact dimensions are large this may have consequences for
 the hierarchy problem. The brane scenario made the notion
  of large compact dimensions, initially put
forward much earlier by Antoniadis in the pre-brane era of string
theory in the context of supersymmetry breaking, more plausible.
It was pointed out in \cite{a90} that the compact dimensions do
not necessarily have to be of Planck scale, and that models can be
constructed that allow even TeV sized dimensions. A different
class of models, involving an alternative, warped, factorization
of the higher dimensional spacetime into a fourdimensional and a
smaller compact part was considered in a simple context by Randall
and Sundrum \cite{rs99a,rs99b}.

The realization that extra dimensions might be large compared to the
Planck scale has lead to a renaissance in model building. In the
present review the focus will be mostly on the interplay between
purely phenomenological models and string theory inspired
constructions which aim at the embedding of the resulting new ideas
into full-fledged fundamental theories, thus completing the cycle.
It will become clear in the process that essential progress has been
made over the past few years in the development of new string
technology. As a result string theoretic and string inspired model
building has reached the point where it is constrained by precision
measurements. String theory thus has become an experimental science.

\vskip .3truein

\section{String theory}

\subsection{Theoretical aspects}

Several excellent introductions to string theory have been
published over the past twenty years, starting with
Green-Schwarz-Witten \cite{gsw}, and continued by  Polchinski
\cite{joe98}. An updated summary of the most recent developments
has appeared in the volume by Becker-Becker-Schwarz \cite{bbs07},
and a more phenomenologically oriented review has been published
by Dine \cite{d07}. In its simplest formulation string theory can
still be viewed very roughly as the description of 1$-$dimensional
objects, open or closed, embedded in some ambient spacetime,
although higher dimensional objects play an important role.

The richness of string theory derives for the most part from its
most dramatic prediction, the fact that a consistent quantum
theory of strings implies the existence of new dimensions or, put
differently, new degrees of freedom. One of the early successes of
string theory was the insight that the specific structure of the
low energy fourdimensional particle spectrum as well as their
couplings are linked to the detailed fine structure of the
geometry of the compact dimensions. One implication of this
relation is an unexpected economy, in that very rough
phenomenological input, e.g. the observed number of generations,
leads to severe restrictions for experimentally viable string
solutions, excluding e.g. most of the Calabi-Yau solutions found
 early on \cite{cdls88, cls90}. The link between the field
 theoretic content of the
models to the specific geometry of the compact dimensions
therefore provides a unification, and also an explanation, of the
fourdimensional spectrum. While the details of the spectrum
therefore clearly must depend on the specific starting point of
the theory, certain of the characteristic features are universal,
leading to model independent predictions. Among these are the
already mentioned existence of additional dimensions, and the
model independent prediction of new particles, such as the
dilaton.

The idea of extra dimensions is of course an old one, formulated
in the context of electromagnetism and gravity by Kaluza
\cite{k23} and Klein \cite{ok}, and revived after a long period of
dormancy around 1980 in the context of supergravity theories. In
supergravity the introduction of additional dimensions was mainly
a tool of convenience, because formulating the various models in
higher dimensions made the derivation of the fourdimensional
actions and their symmetries much more manageable. String theory
is different from these previous attempts in that the extra
dimensions are a necessity, not a convenience.

A second source of complexity arises from the fact that the
worldsheet, swept out by the string as it propagates through
spacetime, supports a rich spectrum of different fields. This, of
course, is not independent of the structure of the ambient
spacetime, but it has not been clear in the past how to construct
a precise relation between the physics on the worldsheet and the
geometry of spacetime. The picture of spacetime as a given, if
dynamical, input must be considered as preliminary. Since string
theory is to be viewed as a fundamental theory it should
eventually lead to a construction of space and time purely in
terms of the
 physics on the worldsheet. Progress in this direction has
 recently been made via the theory of arithmetic modular forms,
  associated to both, the conformal field theory on the worldsheet
  on the one hand, and the arithmetic geometry
 of the compact dimensions on the other \cite{s07}.

 In the mid 1990s it became clear through the discovery
 of several string theoretic dualities \cite{ht95, w95}, in combination
 with Polchinski's D-branes \cite{p95} that,
 although the theory is conceptually simpler than previously thought,
 it is also structurally much richer than
 expected, and that higher dimensional structures play an essential role.
 The simplicity arises from the fact that the string dualities relate
 apparently distinct string theories as being duals of each other.
 In this way quite different looking theories, such as the heterotic,
 type IIA and IIB, type I and I' superstrings, as well as M-theory
 and F-theory,
 emerge as limiting formulations of a single underlying theory.
 The picture of a universal moduli space connecting all the known theories as
 limits of a fundamental one is illustrated in Fig. 1 \cite{p07}.

 \vskip -.1truein
 \hskip 1.3truein
 \includegraphics[height=8cm]{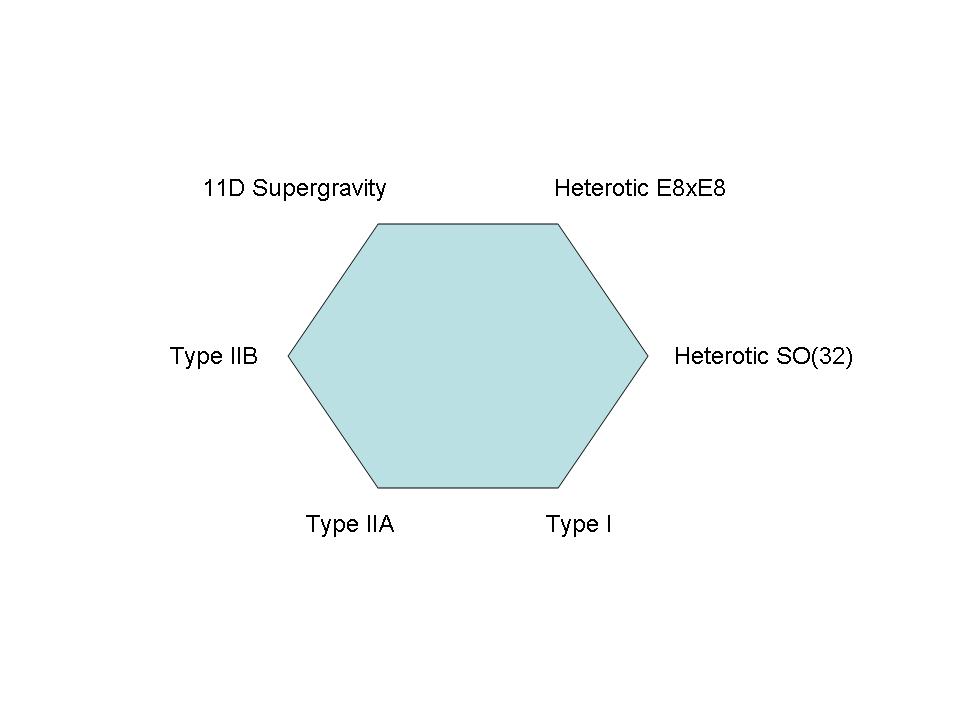}
 \vskip -.8truein
 \begin{figure}[htp]
 \caption{
   An illustration of the universal moduli space that is
 expected to link all known string theories.}
 \end{figure}

 The added complexity introduced into the theory by D-branes
  provides, on the other hand, an enhanced tool box that not only allows to probe the theory
  in deeper ways but, more importantly,  makes it possible to
  envision completely new constructions that were thought
  impossible during the early phase of string phenomenology.
  Furthermore, the introduction of D-branes and their generalizations
   has not only changed
  the face of string theory itself, but it has opened an extensive playing
 ground for many different kinds of ad-hoc models which allow to test a number
 of different ideas in simple ways, such as the notion of large extra dimensions.
 String phenomenology in its early phase adopted the then natural view that
 the compact dimensions were at scales close to the Planck scale,
and that all fields could propagate in all dimensions. Branes on
the other hand can be used to confine fields selectively, and this
opens the way to adjust the size of the compact dimensions to a
dramatic extent. This makes it possible to realize ideas that had
already been used in the pre-brane era of string theory, such as
TeV dimensions \cite{a90} and warped compactifications \cite{s86},
and to construct purely phenomenological models without any
reference to a fundamental theory \cite{add98, rs99a, rs99b}. The
main motivation for these ad-hoc models has been to provide new
ways to think about the hierarchy problem, allowing to translate a
puzzle about energy scales into a geometric puzzle of a disparity
of volumes.

In the current phase of string phenomenology much work is focused
on the embedding of these ideas back into string theory in the
context of different types of D-brane and flux enhanced
compactification schemes. It is in this context that experimental
results ranging from table top experiments to astrophysical probes
are beginning to put constraints on string theoretic and string
inspired models.

\vskip .3truein

 \subsection{Experimental constraints}

Experimental techniques that allow to constrain string theoretic
constructions also impact purely phenomenological models with
similar features, such as extra dimensions, additional fields etc.
The following remarks are concerned with some of the generic
predictions of string theory. Further experimental aspects will
 be discussed in the next section.

\subsubsection{Extra dimensions}

 One of the basic tools in constraining new physics have been
tests of Newton's inverse-square law. This law, earlier envisioned
by Hooke, receives corrections from large extra dimensions and the
propagation of other scalar and vector fields. The new theoretical
developments in string theory, as well as their phenomenological
descendants over the past decade, have renewed interest in
experiments searching  for such deviations from Newtonian gravity
\cite{lcp99, smullin-etal05, adelberger-etal06, kapner-etal06}. In
particular the idea that large compact dimensions might play a
prominent role in string theory, supersymmetry breaking, as well
as for the understanding of the hierarchy problem, has motivated
new attempts to test Newtonian gravity at small distances, where
the existing constraints a decade ago have been rather weak. A
review describing the status of these earlier experiments is given
in \cite{lcp99}, while an extended review of the more recent
efforts that have gone into improving the constraints for such
deviations can be found in \cite{ahn03}.

Complementary early experiments, performed by the Irvine group
\cite{hoskins-etal85}, two Russian groups \cite{dal56, mp88}, and
more recently by Lamoureaux \cite{l97}, have been analyzed in some
detail in \cite{lcp99}, covering between them distance scales
between 1cm and 1$\mu$m. In the recent past several groups have
pursued table top gravity experiments, among them
 the Washington group
   \cite{hoyle-etal01, ahn03, hoyle-etal04, adelberger-etal06, kapner-etal06},
 the Colorado group
    \cite{lcp99, long-etal03},
 the Stanford group
     \cite{chiaverini-etal03, smullin-etal05}, and
 the Purdue group
     \cite{fischbach-etal01, decca-etal05, decca-etal07}.

The deviations from the inverse square law (ISL)  are often
parametrized by a Yukawa type potential
 $$
 U(r) = - G_N \frac{m_1m_2}{r}\left(1 +
        \a \exp\left(-r/\l\right)\right),
 $$
 with a priori free parameters $\a$ and $\l$ parametrizing the strength
 of the new interaction relative to gravity, and the Compton wavelength,
 respectively.
A summary of the current constraints for extra
 dimensions is given in Fig. 2 \cite{kapner-etal06}.
\begin{center}
\includegraphics[width=9cm]{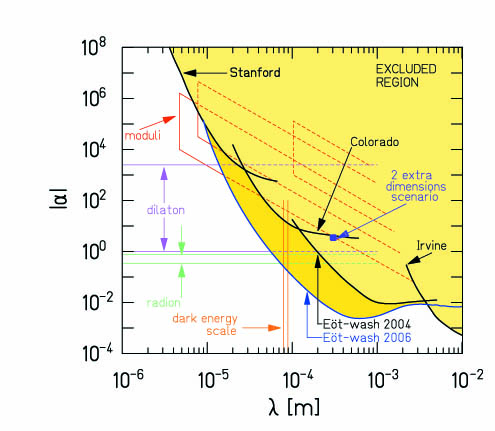}
\end{center}
\begin{figure}[htp]
\caption{
 Constraints on ISL-violating Yukawa interactions with
      $1 \mu {\rm m} < \l < 1 {\rm cm}$. E\"ot-wash 2004 refers to the results
       in \cite{hoyle-etal04}, E\"ot-wash2006 summarizes work done in
       \cite{kapner-etal06}, while the Stanford, Colorado and Irvine results
       are from \cite{chiaverini-etal03,  long-etal03, hoskins-etal85, smullin-etal05}.}
\end{figure}
 The constraints for the strength of possible new interactions
 depend dramatically on the distance. While at a length scale of
 0.1mm additional forces with a tenth of the strength of gravity
 are allowed, at $1/100$ mm the allowed strength is already
 $10^5$ times that of gravity. Extending these experiments to shorter
 distance scales involve Casimir type
 experiments, which test higher $\a-$ranges,
hence they are more relevant for ad-hoc models, which are less
constrained than string theory. Current results for these searches
will be summarized below in the discussion of purely
phenomenological models.

\subsubsection{Constraints on the dilaton mass}

 Precision measurements of Newtonian gravity also lead to
  constraints on the dilaton, a
 scalar partner of the graviton. For the exchange of scalar bosons with mass
 $m$ between two non-relativistic fermions generically leads to
 a potential $U(r) =- g_1g_2 \exp(-r/\l)/4\pi r$, with Compton wavelength
 $\l$. The analyses of the coupling
 of the dilaton to strongly interacting matter shows that the
 coefficient $\a$ for the dilaton exchange is constrained as
  $1\leq \a \leq 1000$ \cite{kw00}.
 If follows from this result that at the 95\% confidence level the
 bound on the dilaton mass is given by \cite{adelberger-etal06}
   $$mc^2 \geq 3.5\rmmeV.$$

\subsubsection{Moduli}

A generic feature of superstring theory, and more generally
supersymmetric models, is the appearance of scalar particles
associated to moduli, fields that parametrize the size and the shape
of the compact dimensions. They couple with gravitational strength
and are massless to all orders in perturbation theory. It may happen
that they become very heavy via nonperturbative effects, in which
case they are not relevant to low-energy phenomenology. If they
remain massless nonperturbatively their mass is determined by the
supersymmtry breaking scale $F$, i.e. $m_{\rmmod} \sim F/M_p$, where
$M_p = 2.4 \times 10^{18} \rmGeV$ is the reduced Planck mass. In
gravity mediated theories these moduli have weak scale masses and
are gravitationally coupled, hence they are not relevant to
phenomenology. In gauge-mediated theories the moduli can be quite
light if the scale of supersymmetry breaking is $(10 \rmTeV)^2$,
leading to Compton wavelengths that are macroscopic \cite{dg96}. In
theories with weak-scale compactification the moduli can have
Compton wavelengths of the order of a millimeter. If the scale of
supersymmetry breaking ranges from $F=1 \rmTeV^2$ to (10 TeV)$^2$
the range for the wavelengths is from $\sim 1$mm to $10 \mu$m
 \cite{add98b, fkz94}. Since the moduli are gravitationally
coupled their existence would lead to deviations in Newton's
inverse-square law at these distances.

Radion exchange e.g. will produce a Yukawa force of the type above
 \cite{cgrt00, ablm03}
 with a strength and a range given by
 \cite{ahn03, hoyle-etal04, adelberger-etal06}
 \beq
  \a = \frac{n}{n+2}
  \eeq
  and
  \beq
   \l \sim \sqrt{\frac{\hbar^3}{cGM^4}} \sim ~ 2.4 \left[\frac{1
   \rmTeV}{Mc^2}\right]^2 \rmmm,
  \eeq
  where $M$ is the unification mass.
 In many cases the radion-mediated force is the longest range
 effect of dimensions because it does not diminish as the number
 of new dimensions increases \cite{add98}.
For $n=1$ and $n=6$ the data of \cite{kapner-etal06} gives
 \bea
  M(n=1) &\geq & 5.7~ \rmTeV/c^2 \nn \\
  M(n=6) &\geq & 6.4~ \rmTeV/c^2.
 \eea
Fig. 2 above shows experimental constraints for moduli, among
other modes.

\subsection{Cosmological constraints}

Cosmological models based on string compactifications are
constrained by recent astrophysical data, in particular the
three-year results for the cosmic background radiation obtained
from the three-year data of WMAP. Of particular interest in this
context are the values of the scalar spectral index $n_s$, defined
via the power spectrum $P_s$ as $n_s = 1 + d\ln P_s/d\ln k$. The
best fit for CMB only is given by $n_s = 0.96 \pm 0.017$, while
 the combination of CMB with large scale structure leads to a
 slightly lower result $n_s = 0.958 \pm 0.015$.

A further parameter that has been the focus of some attention is
the ratio $r$ of the tensor to scalar perturbations. Although the
current limits on the gravity wave contribution $P_t/P_s <0.6$ at
95\% confidence level with CMB data alone is much larger than what
is typically encountered in string inflationary models, the actual
discovery of a gravitation contribution would provide an extremely
important constraint for string models.

\vskip .3truein

\section{Brane models with large dimensions}

A fundamentally new point of view of string theory was suggested
by Polchinski's introduction of D-branes as regions in spacetime
on which physical modes associated to open strings are confined
\cite{p95}. These higher dimensional objects have turned out to be
important in particular for the many string theoretic dualities
that suggest a unified picture of string theory. For example,
mirror symmetry \cite{cls90, gp90} emerges as
 a kind of T-duality \cite{syz96, hv00}, which in turn has implications
 for the construction of spacetime from the string worldsheet
 \cite{s07}.

\subsection{Ad hoc models}

The D-brane idea has transcended string theory.  One purely
phenomenological application that has had an impact on current
string model building is an attempt to understand the hierarchy
problem by combining ad-hoc brane scenarios with an earlier, at
the time provocative, idea that arose in the context of string
theory. Antoniadis suggested in \cite{a90} the idea that the
existence of large dimensions, possibly as large as 1 TeV, would
provide a strategy to address the difficult issue of supersymmetry
breaking in string theory (see also \cite{l96}). Arkani-Hamed,
Dimopoulos and Dvali (ADD) \cite{add98} later observed that if
such large dimensions exist they have implications for our
thinking about the hierarchy problem, i.e. the puzzle that the
weak scale in four dimensions $\sim 250 \rmGeV$ is very much lower
than the Planck scale in four dimensions $\sim 10^{19} \rmGeV$, a
huge step for which one would like to have an explanation.

In the context of large compact dimensions the problem of a large
hierarchy can be traded for the problem of explaining a large
volume, since the four-dimensional Planck scale $M_p=\sqrt{\hbar
c/G}$ is related via the effective action to the fundamental (bulk
theory) scale $M_f$ as
 $$M_p^2 = M_f^{n+2}V^n.$$
 Here the bulk theory is assumed to have $(n+4)$ dimensions and the
Lorentzian spacetime $M^{1,n+3}$ splits into a fourdimensional
part and a compact manifold $K^n$ as $M^{1,n+3}=M^{1,3} \times
K^n$, and units with $c=1=\hbar$ have been used. The basic
observation now is that the fundamental scale in the bulk could be
much lower than the fourdimensional Planck scale because the two
are linked via the volume $V^n$ of the compact manifold $K^n$. If
the size of these extra dimensions is large this would explain the
huge value of the Planck scale in terms of a relatively low
fundamental scale $M_f$.

\subsection{Experimental constraints}

In the simplest possible compactification scheme the realization
of the brane picture formulated by ADD involves only circles,
leading to a toroidal structure of the extra-dimensional part of
spacetime.  Assuming that all radii are the same introduces only a
single new scale, the radius $R$ of the circles. It is determined
by the number $n$ of dimensions considered,
 $R =
 (M_p/M_f)^{2/n}(1/M_f)$, and experimental data provides
constraints on the number of possible dimensions. If $M_f$ is
chosen to be $1 \rmTeV$ the case $n=1$ would be readily excluded
because this model leads to an astronomically sized fifth
dimension. Increasing $n$ within the limits of interest to string
theory shows that they are all unacceptably large because particle
accelerators should have seen them. This is where the D-brane
picture enters. Assuming that all the fields of the standard
models are confined to a D3-brane leaves only gravity as a
possible probe of the compact dimensions.

Fig. 2 above shows that the constraints provided by gravity are
surprisingly weak, leaving a wide open window for new interactions
at large scales. This observation, made already in \cite{dg96},
 not only triggered new interest in experimental tests of Newton's law on
small scales, as pointed out already, but also in phenomenological
analyses of possible bounds on the string scale from collider
processes \cite{hlz99, h99}. In the context of phenomenological
models a wider range of possible interaction strengths is of
interest than indicated in Fig. 2. Data constraining new
interactions on shorter distance scales are shown in Fig. 3
\cite{ahn03}, which is adapted from \cite{fischbach-etal01}.
 \begin{center}
\includegraphics[width=8cm]{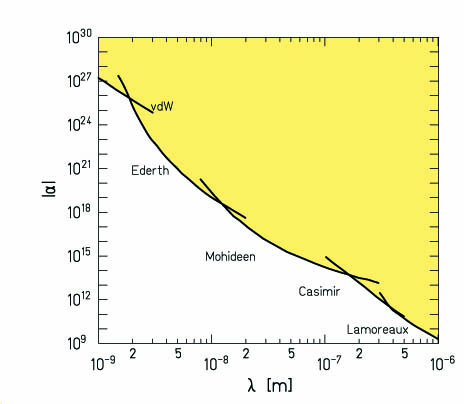}
\end{center}
\begin{figure}[htp]
\caption{\small
 Constraints on ISL-violating interactions, summarized from data in  with
      $1 {\rm nm} < \l < 1 \mu{\rm m}$.}
\end{figure}

The individual constraints in the diagram are based on the data from
different experiments, notably Lamoreaux \cite{l97}, Casimir
\cite{dal56}, Mohideen \cite{hcm00}, Ederth \cite{e00} and van der
Waals (vdW) \cite{it72}. The theory underlying experiments measuring
Casimir forces is much more involved, because they deal with
surfaces of real metals, and several corrections have to be taken
into account, among them the roughness of the surfaces, as well as
finite temperature corrections. The importance of these corrections,
and their impact on constraining new physics, has been
controversial. A review of the current status of these issues is
provided in \cite{mostepanenko-etal05}. Long distance constraints,
finally, are summarized in Fig. 4.
 \begin{center}
\includegraphics[width=9cm]{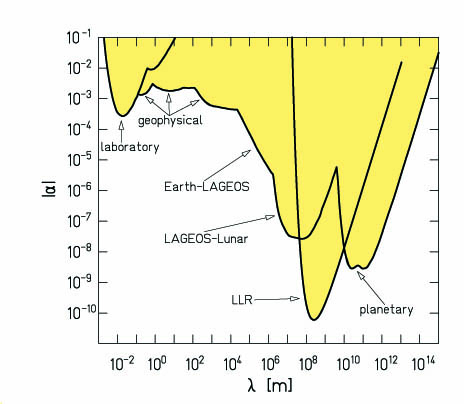}
\end{center}
\begin{figure}[htp]
\caption{\small
 Constraints on ISL-violating interactions, summarized from data in  with
      $1 {\rm mm} < \l < 10^{15} {\rm m}$.}
\end{figure}

The assumption of ADD type models that all fields except the
graviton propagate on a three-dimensional brane, and only the
graviton propagates in the bulk, implies that new contributions
   to collider processes arise from the Kaluza-Klein excitations of the
   graviton. Although the
   contributions of individual Kaluza-Klein modes, with 4D gravitational
   strength, to collider processes is extremely small, a very
   large number of such modes contribute in a TeV-scale collider
   process because the compactification scale is so small.
   The net Kaluza-Klein effect can cause a significant deviation
   from the standard model production
   rates. Bounds on the string scale from analyses of various
   collider processes are typically of the order of a TeV for
   these symmetric compactification models \cite{st98, mpp99, grw99,
   h99}.

\subsection{Embedding of the D-brane scenario into string theory}

In order to make the scenario considered in \cite{add98} into a
consistent physical picture it is necessary to derive this class
of models from string theory. It was outlined in \cite{aadd98}
that such an embedding can be achieved by considering the
so-called type I theory, a model which contains open and closed
strings \cite{pw96}. The realization considered in \cite{aadd98}
involves the localization of the standard model fields to a
D3-brane, while gravity also propagates in the bulk.

More generally, one can envision scenarios where the standard
model can be viewed as localized on a $p-$dimensional brane with
$p\geq 3$ while closed strings can move in the bulk. In the
simplest compactification schemes this would give rise to the
production of massive standard model Kaluza-Klein towers in
colliders once the energies reach the compactification scale of
the 'longitudinal' $(p-3)-$dimensions. The characteristic signal
in particle colliders is graviton emission into the bulk, leading
to missing energy that escapes detection.

\vskip .3truein

\section{Universal and mixed extra dimensions}

\subsection{UEDs}

 The original assumption in the pre-D-brane period of
string theory was that all fields propagate in the compact
dimensions. This notion has since been renamed in the context of
phenomenological ad-hoc models with large dimensions as the
so-called universal extra dimensions (UED) scenario, in
distinction to D-brane models. In this case momentum conservation
in the extra dimensions implies the conservation of Kaluza-Klein
mode number at tree level. This controls all the couplings of
these fields, and each vertex involves at least two such
excitations, hence direct production is only possible in pairs of
Kaluza-Klein states, and there are no tree-level contributions to
the electroweak observables. The main constraint observed in
\cite{acd01}
 comes from weak-isospin violation effects. The mass bounds for
 the first excited modes are relatively low \cite{acd01, ddgr00, r01}.
 For one extra dimension the limits on $R^{-1}$ are between 300 GeV and 500 GeV,
 depending on the Higgs mass \cite{ay03}. This opens the
 possibility of discovering Kaluza-Klein states in this class of
 models in upcoming collider experiments.

 The fact that in UED models Kaluza-Klein-parity conservation leads
 to a stable lightest Kaluza-Klein state has the consequence that in these models
 a lightest state of this type could provide a dark matter candidate. It can constitute
 cold dark matter if its mass is approximately 600 GeV \cite{st03,
 bk05, km06}, well above current collider limits. The LHC might
 be able to test these models up to $R^{-1} \sim 1.5 \rmTeV$
 \cite{cms02}.

\subsection{Mixed extra dimensions}

The 'pure' scenarios of fixing the standard model on a D3-brane,
or allowing them to propagate into all compact dimensions in the
UED models, raises the obvious question whether phenomenological
constraints change in mixed models, in which some of the fields
are confined to the brane and others are allowed to propagate in
the bulk. Non-universal models in which the gauge bosons propagate
in the extra dimensions, but the fermions are confined to the
standard model D3-brane \cite{aab00, dolzh00}, do provide
different collider bounds than completely confined models. These
include effects on electroweak precision measurements \cite{rw00},
Drell-Yan processes in hadronic colliders \cite{mp99, nyy99},
$\mu^+\mu^-$ pair production in electron-positron colliders
\cite{mp99, nyy99}, electroweak processes in high energy
electron-positron colliders, as well as multijet production in
hadronic colliders \cite{dmn02}.

\vskip .3truein

\section{Warped compactifications}

\subsection{Randall-Sundrum models}

An alternative compactification scheme in the context of producing
phenomenological models with large hierarchies involves the
warping of the four-dimensional part of spacetime of the compact
dimensions. Traditionally the ansatz for a metric in a
Kaluza-Klein or string theory framework on a $D-$dimensional
spacetime for spacetime with coordinates $(x^{\mu},y^m),
\mu=0,...,3, m=1,...,D-4$ has been a simple product
 $$
 ds^2 = \sum_{\mu \nu} g_{\mu \nu}(x^{\k}) dx^{\mu}dx^{\nu} +
 \sum_{mn}g_{mn}(y^k)dy^mdy^n,
 $$
 where the $x^{\mu}$ are the standard fourdimensional coordinates
 and the $y^m$ parametrize the compact dimensions.
In the warped case generalized solutions of the equations of
motion are considered for which the normalization of the
four-dimensional metric varies in the transverse direction
 $$
 ds^2 = e^{A(y^k)}\sum_{\mu \nu} g_{\mu \nu}(x^{\k}) dx^{\mu}dx^{\nu} +
 e^{-A(y^k)} \sum_{mn}g_{mn}(y^k)dy^mdy^n.
 $$
 In
this scenario a given invariant energy scale can give rise to many
four-dimensional scales, depending on the position-dependent
gravitational redshift in the transverse space. This mechanism has
played an important role in particular in the Randall-Sundrum
models \cite{rs99a, rs99b}. In the compact version \cite{rs99a} a
fivedimensional spacetime is considered with a metric of the form
 $$
 ds^2 = e^{-2kr_c|\phi|}\eta_{\mu \nu}dx^{\mu \nu}dx^{\mu}
 dx^{\nu} + r_c^2 d\phi^2,
 $$
 where $x^{\mu}$ are the fourdimensional Minkowski
 space coordinates, $-\pi \leq \phi \leq \pi$, with
 identification of $(x,\phi)$ and $(x,-\phi)$, and $r_c$ sets
 the compactification scale. Two 3-branes are
 located at $\phi=0$ and $\phi=\pi$ respectively.
 This set-up is similar to a compactified version of the Horava-Witten
 scenario where M-theory is compactified on Calabi-Yau manifolds
 \cite{hw96a, hw96b, w96}.

The generation of a hierarchy via redshift has a number of
interesting consequences. It can be shown that the
 effective fourdimensional Planck scale $M_p$ is given by
 $$ M_p^2 = \frac{M^3}{k}(1- e^{-2kr_c\pi}).$$
 This implies that even at large $kr_c$ the Planck scale is of the
 same order as the fundamental scale.
 A further implication
  of the warp factor in the metric is that a field confined to the 3-brane at
$\phi=\pi$ with mass parameter $m_0$ will have a physical mass
given by $m=m_0 e^{-kr_c\pi}$. For $kr_c$ about 12 this implies
that the weak scale therefore is dynamically generated on the
visible brane from the fundamental Planck scale. Furthermore,
thresholds to the production of Kaluza-Klein modes can be reached
at low energies, in the TeV range, and
 scattering at low energies can reach the fundamental
 Planck scale, due to the relative redshift.

 These results also suggest the possibility of experimental
 probes of Planck- or string scale physics at energies far below
 the apparent four-dimensional Planck scale. As an example the
 possibility has been suggested in \cite{adm-r98, ehm00}
 that black holes could be produced at relatively low energy scales,
 leading to the idea that the CERN Large Hadron Collider could become a black hole
 factory. Such TeV black holes radiate mainly on the brane \cite{ehm00}, and
 detailed aspects of possible LHC signatures have been
 discussed in \cite{gk01, gt02, dl01}.

\subsection{Embedding of warped models into string theory}

Warped metrics have been considered in string theory early on by
Strominger \cite{s86}, who observed that fluxes can be considered
if the metric is generalized from a pure product to a twisted
form. They have been revived in the context of M-theory by
Becker-Becker \cite{bb96} and in the framework of F-theory in
\cite{drs99}.  A realization of the compact Randall-Sundrum
scenario within string theory was first discussed in detail by
Giddings-Kachru-Polchinski (GKP) in \cite{gkp02}, developing
further a construction by Verlinde of the warp in terms of $N$
coincident D3-branes on a Calabi-Yau manifold \cite{v00}. The
particular importance of the GKP discussion of warped
compactifications in string theory derives from their observation
that it is possible in this framework to fix in several classes of
string models many of the large number of moduli usually
associated to supersymmetric vacua. This represents important
progress from a practical point of view, because the stabilization
of these moduli opens the door to a much more precise
phenomenological analysis of these models than previously
possible.

In order to fix the moduli it is necessary to break conformal
invariance and most of the supersymmetry. The strategy to achieve
the necessary reduction of symmetry is to place $D3-$branes not at
a smooth point of the transverse space, but at a singularity. A
detailed construction involving the introduction of O3-planes,
D7-branes, and fluxes in addition to the D3-branes of Verlinde
within type-IIB solutions has been given by GKP. Concrete examples
are constructed as orientifolds of CY compactifications, and also
as F-theory compactifications. The latter are of interest because
they allow larger fluxes and hierarchies. In many cases the
resulting solutions stabilize the dilaton and the complex
structure moduli at a high scale, and further corrections to the
leading order action are needed to generate a potential for the
remaining ones.

The GKP construction of \cite{gkp02} was enhanced by
 Kachru-Kallosh-Linde-Trivedi (KKLT) \cite{kklt03} by taking nonperturbative
 corrections to the superpotential into account, and by adding
 anti-D3-branes into the mix as a further ingredient. The combination
 of these two ingredients allows not only to stabilize the remaining (K\"ahler)
 moduli, but also to lift the resulting anti-de Sitter space to a solution
 with a positive cosmological constant.
 By tuning the flux superpotential to be very small it is possible
 to obtain gravitino masses of $\cO(1-10\rmTeV)$. This has triggered much work on the
distribution of flux vacua, testing whether one can achieve the
required small values necessary for supersymmetric vacua at large
volume \cite{ad04, dd04, gktt04}.

A somewhat different approach to realize the warping idea within
string theory has been pursued in the 'large volume
compactification' approach, introduced in \cite{bbcq05}, and
developed further in \cite{cqs05, aqs05, cq06}. It is based on the
observation \cite{bb04} that the inclusion of string corrections
to the tree-level supergravity effective action, previously
computed in \cite{bbhl02}, leads to interesting modifications of
the models considered by GKP and KKLT. This class of vacua
preserves the main ingredients of the GKP-KKLT scenario as far as
the orientifold and D-brane structure within type IIB theory is
concerned, leading again to a stabilization of all the moduli, but
allowing a limit with a large overall volume modulus  and all the
remaining moduli small. Fluxes stabilize the dilaton and the
complex structure at a high scale, while the flux superpotential
does not have to be fine tuned to small.

It is argued in \cite{bb04} that the maximum value that the flux
induced superpotential can achieve scales as the square root of
the Euler number of the Calabi-Yau fourfold giving rise to the
type IIB model considered. The construction in \cite{lsw99} of the
class of about $10^6$ Calabi-Yau fourfolds derived from
Landau-Ginzburg potentials shows that these values can easily be
of order $10^3$. Masses derived for the gravitino for a specific
compactification manifold lead to intermediate scales, and are
independent of the fluxes. Thus this framework provides a
realization of the intermediate scale string scenario considered
in \cite{b99, biq99}. Inflationary models based on K\"ahler moduli
considered in \cite{cq05} lead to spectral indices in the $1-2/N_e
\sim 0.96$ range, where $N_e$ is the number of e-foldings. This
result is consistent with the three-year WMAP results
\cite{spergel-etal06}. The tensor-to-scalar ratio is extremely
low, much smaller than the current experimental limit of about
$0.3$. An observation of tensor modes would therefore invalidate
this class of models, along with many others \cite{k07}. An
extension of the framework of large volume compactification to
incorporate the effects of strong warping has been discussed in
\cite{burgess-etal06}, while the inclusion of further corrections
has been pursued in \cite{bhp07}.

\vskip .3truein

\section{Concluding remarks}

 Many different interesting (classes of) string theoretic, or
string inspired, models exist that could not be described in any
detail in this brief summary. The focus here is on the general
structure of string theoretic models, as well as on ideas that
 are on the interface between string theory in particular, and particle physics
 in general. The models reviewed are not only of interest merely as ad hoc
 models, but they emerge as limits of
 fundamental, string theoretic, solutions.

 Some of the prominent models, such as the KKLT model \cite{kklt03}, and
 the large volume model (see e.g. \cite{bbcq05, cqs05}), are analyzed in
 greater detail in ref. \cite{kks06}. This paper addresses the
interesting 'inverse problem' in the context of the LHC: it treats
the issue whether phenomenological data from the LHC can in
principle be used to distinguish between the large number of
different types of string models that have been proposed to date.
Among those considered are not just the KKLT and the large volume
models, but also a number of heterotic constructions. A discussion
of the issues introduced by the fine-tuning of the cosmological
constant can be found in \cite{sda07}.

 A subject that could barely
be mentioned here, and that deserves its own review, is that of
the implications of the recently considered classes of warped flux
compactification models for early cosmology. An update of these
developments can be found in \cite{q03, b06, k07}.

The great advances in string technology that have been achieved
over the past few years, in combination with new experiments, such
as much more precise table top gravity tests, the WMAP probe, and
the upcoming LHC, have transformed string theory as a discipline.
They have triggered a renaissance in string phenomenology, leading
to a continuous spectrum of ideas that spans the range from
full-fledged string models to much simpler and easily accessible
ad-hoc constructions. They have also done much to strengthen the
links between experimenters, particle physics phenomenologists,
and string theory model builders.

\vskip .3truein

{\large {\bf Acknowledgement.}} \hfill \break
 It is a pleasure to thank Monika Lynker for discussions and Ilka
 Brunner, Shanta de Alwis, Wolfgang Lerche, Rob Myers, and Fernando Quevedo
 for correspondence. This work was supported in part by an IUSB
 Faculty Research Grant.

 \vskip .3truein

\section{Glossary}

{\bf Compactification.} \hfill \break
 The notion that a higher dimensional spacetime factors into a
 the standard observable universe of $(3+1)$ dimensions plus an
 additional number of compact dimensions of finite volume. In the
 context of the standard big bang model this is a misnomer $-$ it
 would be more appropriate to think of a dynamical process that
 decompactifies some dimensions while keeping others small. The
 first idea in this direction came from Brandenberger and Vafa
 \cite{bv88}, and has more recently been pursued in
  \cite{brian}.

\vskip .15truein

{\bf D-branes.} \hfill \break
 In any theory with open strings a D-brane is a region of
 spacetime on which open strings can end. The name derives from
 the choice of Dirichlet boundary condition at the ends of the
 open string, as opposed to Neumann conditions.

\vskip .15truein

{\bf Fluxes.} \hfill \break
 In string theory in general, and type II and F-theory in particular,
  there are several tensor fields that generalize the vector
  potential of electromagnetism. The
 associated field strengths lead to fluxes that are constrained by
 a tadpole cancellation condition.

 \vskip .15truein

{\bf Hierarchy problem.} \hfill \break
 The idea that the small number given by the ratio of the weak
 scale and the Planck scale should have an explanation in terms of
 some fundamental aspects of a unified theory.

 \vskip .15truein

{\bf Kaluza-Klein modes.} \hfill \break
 In a compactification scheme with periodic boundary conditions
 in some dimensions any field can be expanded in a Fourier series
 along these compact directions. This results in an infinite tower
 of massive states whose masses are determined by the winding
 number and the length scale associated with the compact
 dimension.

 \vskip .15truein

{\bf Large extra dimensions.} \hfill \break
 Any dimension much larger than the Planck scale is considered to
 be large.

\vskip .15truein

 {\bf Moduli.} \hfill \break
  Parameters that describe the shape of the compact
  manifold which appears as a factor of the higher dimensional
  manifold in string theory. In terms of the underlying conformal
  field theory moduli are marginal operators with no potential to
  all orders in perturbation theory. Two prominent classes of
  moduli associated to Calabi-Yau manifolds are the moduli associated
  to the K\"ahler deformations, determined by the (K\"ahler) metric
  on the manifold, and the moduli associated to complex
  deformations.

\vskip .4truein

\parskip=0.001truein
\baselineskip=13pt

\begin{small}

\end{small}

\vskip .3truein

\baselineskip=21.5pt

{\bf Further reading.} \hfill \break
 A comprehensive introduction to the developments in string theory
 in the past decade can be found in \cite{bbs07}. A more
 phenomenologically oriented discussion can be found in Dine's
 review \cite{d07}.
 An elementary introduction to the basics of lower dimensional
 Kaluza-Klein theory is given in Sundrum's lectures \cite{s05}.
 More recent developments along the lines of incorporating the
 warped dimensions idea into string theory have been reviewed
 extensively in \cite{g06, dk06, bkls06}.
 A review on black hole aspects at the LHC for TeV quantum gravity
 can be found in \cite{l06}.
 An extended introduction to the potential aspects of string phenomenology at
 the LHC is given in \cite{kks06}.

\end{document}